\documentclass[12pt,a4paper]{article}
\newcommand\be{\begin{equation}}
\newcommand\ee{\end{equation}}
\newcommand\ket[1]{|#1\rangle}
\newcommand\bra[1]{\langle #1|}

\begin{document}
\title{Reconstruction of spin states and its \\ conceptual implications\footnote{Contribution to the proceedings of the conference {\sl New Insights in Quantum Mechanics}, 9/98 D-Goslar (to appear)}}
\author{Stefan Weigert\\
Institut de Physique, Universit\'e de Neuch\^atel\\
Rue A.-L. Breguet 1, CH-2000 Neuch\^atel, Switzerland\\
\texttt{stefan.weigert@iph.unine.ch}}
\date{September 1998}
\maketitle
\begin{abstract}
State reconstruction for quantum spins is reviewed. Emphasis is 
on nontomographic approaches which are based on measurements performed with a Stern-Gerlach 
apparatus. Two consequences of successfully implemented state reconstruction are pointed out. First, it allows one to determine experimentally the expectation value of an arbitrary operator $\widehat {\cal O}$ without a device measuring it. Second, state reconstruction suggests a reformulation of Schr\"odinger's equation in terms of expectation values only, without explicit reference to a wave function or a density operator. 
\end{abstract}
In a footnote of his article on quantum mechanics for the ``Handbuch der Physik,'' Pauli 
hides a question which, apparently, he does not know to answer \cite{pauli33}:
\begin{quote}
Die mathematische Frage, ob bei gegebenen Funktionen $W(x) $ und  $W(p) $ die Wellenfunktion $\psi$ stets eindeutig bestimmt ist, wenn es eine 
solche zugeh\"orige Wellenfunktion \"uberhaupt gibt, (d. h. wenn $W(x)$ und $W(p)$ physikalisch vereinbar sind), ist noch nicht allgemein untersucht worden.
\end{quote}
Knowledge of the probability distributions $ W(x)\equiv| \psi(x)|^2  $ und  $ W(p) \equiv |\psi(p)|^2  $ for position and momentum is equivalent to the knowledge of all moments $\bra{\psi} \hat{x}^n \ket{\psi}$ and $\bra{\psi} \hat{p}^n \ket{\psi}$, $n = 1, 2, \ldots$ Therefore, one can immediately rephrase Pauli's question in terms of quantities 
which are measurable in experiments, at least in principle. Is it possible to reconstruct the state $\ket{\psi}$ (of a particle in a potential $V(x)$ on the real line, say) on the basis of a collection of expectation values, i.e., 
\begin{equation}
\{ \bra{\psi} \hat{x}^n \ket{\psi}, \, \bra{\psi} \hat{p}^n \ket{\psi}, 
                                                \,  n = 1, 2, \ldots \}
 \Rightarrow \ket{\psi} \, ?
\label{basicquestion}
\end{equation}
As it stands, Pauli's question must be answered in the negative: the {\em Pauli data} on the right-hand-side of (\ref{basicquestion}) do not {\em always} single out a unique state $\ket \psi$. A wave function $\psi(x)= \psi_r(x) + i \psi_i(x)$ with linearly independent real and imaginary parts,
\be 
\alpha \psi_r(x) + \beta \psi_i(x) = 0 \Leftrightarrow \alpha = \beta = 0\, ,
\label{linindep}
\ee
and definite parity,
\be
\psi (x) = - \psi (-x) \, ,
\label{antisym}
\ee
provides a simple counterexample \cite{parity??}. The state $\widetilde{\psi} (x) = \psi^*(x)$, obtained by complex conjugation of the wave function $\psi(x)$, is linearly independent of $\psi(x)$. It represents a {\em Pauli partner} of the original state  since it leads to the same probability distributions:
\be
|\widetilde{\psi}(x)|^2                     = | \psi(x)|^2 \,  \mbox{ and } 
|\widetilde{\psi}(p)|^2 = | - \psi^* (p)|^2 = | \psi(p)|^2 \, .
\label{partnerex}
\ee
According to \cite{corbett+78}, families of {\em Pauli non-unique} states (not necessarily of definite parity) exist which are dense in the one-particle Hilbert space. 

On the other hand, a coherent state $\ket{\alpha}$, defined as eigenstate of the annihilation operator $a$, is easily seen to be defined completely by the probability distributions of position and momentum. In fact, the expectation values of the operators $\hat{x}$ and $\hat{p}$ are already sufficient,
\be
\bra{\alpha} a \ket{\alpha} 
 = \alpha 
 = \sqrt{\frac{m \omega}{2 \hbar}} \bra{\alpha} \hat{x} \ket{\alpha} 
   + \frac{i}{\sqrt{2 m \omega \hbar}} \bra{\alpha} \hat{p} \ket{\alpha} \, ,
\label{coherent}
\ee
since they determine the complex number $\alpha$ and thus the state $\ket{\alpha}$.
\subsection*{General setting of state reconstruction}
Let us rephrase the problem of state reconstruction from a more general point of view. Consider a quantum mechanical system ${\cal S}_{QM}$ described by a Hamiltonian operator $\widehat{H}$ acting in a Hilbert space $\cal H$ of states. The states of the system may be pure ones or mixed ones, in both cases described conveniently by a density matrix $\hat{\rho}$. {\em Given} the state, it is straightforward to predict the expectation values of arbitrary hermitean operators $\widehat{\cal{O}}_a = \widehat{\cal{O}}_a^\dagger$ in the algebra $\cal{A}$ of operators acting on  $\cal H$:
\be
\langle \widehat{\cal O}_a \rangle_{\hat{\rho}} 
   = \mbox{Tr} \left( \hat{\rho} \widehat{\cal O}_a \right)  \, , \quad 
\widehat{\cal O}_a \in {\cal A} \, .
\label{expvalues}
\ee
These numbers are to be directly compared with the outcome of appropriate measurements 
performed on the physical system ${\cal S}_{QM}$. 

In this setting, the problem of state reconstruction is clearly seen to define an {\em inverse} problem. The the state of the system with density matrix $\hat{\rho}$ is unknown, while the values of various expectation values\footnote{For simplicity, the standard formulation of (nonrelativistic) quantum mechanics will be assumed here, including the idealizations  that the preparation of the states to be measured is {\em perfect}, that {\em completely} reliable detectors exist, and that one is able to handle {\em infinite} ensembles in order to extract expectation values. The modifications required for a realistic experimental setup have been studied in \cite{buzek+96}, for example.}
are assumed to be given:
\begin{equation}
\left\{ \langle \widehat{\cal{O}}_j \rangle   \, , \,   j \in J \right\}
               \stackrel{?}{\Rightarrow} \hat{\rho}\, ,
\label{general}
\end{equation}
where the index $j$, taking values in some set $J$, labels hermitean operators. Conceptually, the situation is similar to a scattering problem. Given a scattering 
potential, the calculation of cross sections is straightforward, infering the underlying potential from scattering data, however, is highly complicated.

From the outset it is not obvious which data should be collected experimentally in order 
to reconstruct the state of the quantum system. For example, in the case of a single particle the position and momentum probability distributions as proposed by Pauli
do not completely characterize the unknown pure state. In other words, the 
selection of a set of observables to be measured is by no means trivial. Therefore, one might expect that different solutions to Pauli's problem can be found. 
\begin{itemize}
\item{Suppose that the expectation values of {\em all} hermitean operators $\widehat{\cal{O}} \in \cal{A}$ were known at some instant of time.\footnote{Clearly, this assumption is debatable due to the difficulty of associating a measurement procedure with each hermitean operator -- see, however, the last part of this paper.} This should be 
sufficient information to determine the density matrix $\hat{\rho}$ since all the information one ever would like to calculate from $\hat{\rho}$ is already given.
Nevertheless, even in this case it is not obvious how to actually {\em calculate} the density operator as a function of the measured expectation values.}
\item{The notion of a ``quorum'' ${\cal Q}$  denotes a set of operators 
sufficient to be measured in order to extract the underlying quantum state \cite{band+79}.  As follows from counting the free parameters of the density matrix of a spin $s$
(cf. below), a quorum of operators $ \widehat{\cal O}_q \in {\cal Q}$ will indeed be a {\em subset} $ {\cal Q} \subset \cal{A}$ of all the operators which act on the Hilbert space $\cal{H}$ of the system. An ideal and realistic quorum of operators would consist of the {\em minimal} number of some basic operators for which the 
measurement is experimentally {\em feasible}.}
\item{Instead of measuring many expectation values at a fixed time, one might follow the time evolution of a few observables over some period of time. Such an approach has led to the result \cite{leonhardt+96} that one can reconstruct a particle state in a known one-dimensional potential by following the probability distribution of position, $|\psi(x,t)|^2$, over the interval $-\infty < t < \infty$.}
\item{The problem is highly sensitive to the conditions which are imposed from the beginning. Knowing from the outset that the system under study is in a pure state, one can reduce considerably the data necessary for state reconstruction. Suppose the 
probability distribution, $|\psi(x,t_0)|^2$, measured at time $t_0$ has no nodes for finite values of $x$. Then the state $\ket{\psi}$ is uniquely determined by a second measurement of the position probability distribution at an infinitesimal time $\Delta t$ later, $|\psi (x,t_0+ \Delta t)|^2$  \cite{weigert96}. In the non-generic case of a pure state with $N$ nodes (requiring both the real {\em and} the imaginary part of the wave function to vanish simultaneously at $N$ positions), these data are compatible with a continuous manifold of states isomorphic to an $N$-dimensional torus. This is due to the fact that $N$ nodes 
divide the real line into $N+1$ compartments with wave functions the relative phases of which remain undetermined. A small number of additional expectation values is necessary in order to determine their values. The multi-dimensional version of this result \cite{wallstrom94} casts doubts on the proposed equivalence of Madelung's hydrodynamic formulation of quantum mechanics \cite{madelung26} to Schr\"odinger's formulation.}     
\item{The {\em tomographic} method of state reconstruction turns out to be particularly attractive. This approach is based on two ingredients. In a phase-space formulation of quantum mechanics a quantum state is associated with a unique a quasi-probability or Wigner function \cite{baker58} and {\em vice versa}. Its marginals are accessible experimentally, and an (inverse) Radon transformation \cite{bertrand+87,risken+89} allows one to reconstruct the phase-space distribution. This method has been applied to various physical systems: quantum states of vibrating molecules \cite{dunn+95}, of trapped ions \cite{leibfried+96}, as well as the state of atoms in motion \cite{kurtsiefer+97} have been reconstructed successfully in the laboratory. Similarly, quantum optical experiments \cite{smithey+93} have been performed. Reviews can be found in \cite{raymer97} and \cite{leonhardt97}.}
\end{itemize}  
As can be seen from the general formulation of the problem, it also arises naturally for spin systems. The essential difference to particle systems is its setting in a {\em finite-dimensional} Hilbert space. Therefore, one might expect that the problem of state reconstruction simplifies. From now on, the focus will be on the Pauli problem for a single spin with 
quantum number $s$. 
\subsection*{Reconstructing a mixed spin state}
The Hilbert space of a spin $s$ has $(2s+1)$ complex dimensions. Therefore, the most general (unnormalized) density matrix $\hat{\rho}$ is a $(2s+1) \times (2s+1)$ hermitean matrix with  $(2s+1)^2$ real parameters. Various methods have been proposed to determine $\hat{\rho}$.

The expectations of $4s(s+1)$ linearly independent spin multipoles do fix a unique (normalized) density operator \cite{band+71a}. However, no method is outlined which would indicate how to determine their values experimentally. 

If a {\em Feynman filter} \cite{feynman+65} were available, a phase sensitive version of a Stern-Gerlach apparatus, one could determine directly moduli and (relative) phases of the individual matrix elements of the density operator \cite{gale+68}. It is, however, not obvious whether such an apparatus can be build in the laboratory. 

In order to establish a down-to-earth approach, it is natural to restrict the measurements to those performed with a standard Stern-Gerlach apparatus, the quantization axis of which can be oriented arbitrarily in space. In this spirit, the density matrix of a spin $s$ has been shown to be fixed through $(4s+1)$ measurements using a Stern-Gerlach apparatus \cite{newton+68}. All the directions involved are assumed to be located on a cone about some fixed axis in space. Clearly, this will not be the most efficient method since the number of experimentally determined parameters 
$(=(4s+1)(2s+1))$ exceeds the number of free parameters. Along similar lines, an explicit expression for the density matrix has been derived \cite{amiet+98a} in terms of just $(2s+1)^2$ intensities, that is, 
\be
\hat{\rho} \Leftrightarrow \{ p_m^{(k)}, -s \leq m \leq s, k=1,2,\ldots, 2s+1\} \, ,
\label{mixedrho}
\ee
with $p_m^{(k)}$ being the probability to obtain the value $m$ when the spin is measured along the $k$-th axis out of $(2s+1)$ directions in space. 
The $(2s+1)$ axes are located on a cone about the $z$ axis, for example. This result is satisfactory since it pertains to a minimal quorum of observables all of which are directly accessible in an experiment. 
\subsection*{Reconstructing a pure spin state}
Suppose now that the spin state to be reconstructed is known to be prepared in a (normalized) {\em pure} state which has  $4s$ parameters. This number is {\em linear} in $s$ while the data needed for state reconstruction presented above grows {\em quadratically} with $s$. Especially for large values of $s$, one would like to systematically reduce the amount of data needed to reconstruct a pure state. Two methods are presented which solve this problem in an optimal way with respect to Stern-Gerlach type measurements. 

The first approach involves the measurement of intensities along {\em three} axes, two of which are infinitesimally close to each other, while the third one is perpendicular to the plane spanned by the other two \cite{weigert92}. It is possible to make explicit the set 
of Pauli partners compatible with the intensities associated with the two nearby axes.  Their number is $2^{2s}$, growing thus  exponentially with the spin quantum number $s$. 
Effectively, on has to solve $2s$ {\em quadratic} equations for the $2s$ unknown phases;
this leaves $2s$ signs undetermined giving rise to $2^{2s}$ possible combinations. The third measurement provides $2s$ additional real numbers which can be used to select the correct state among the $2^{2s}$ partners. Hence, a total of $6s$ numbers has to be measured -- clearly, this still exceeds $4s$, the number of parameters of a pure state, but with a Stern-Gerlach apparatus apparently one cannot do better. 

The second appraoch \cite{amiet+98b} is more realistic than the first one because it is not necessary to perform measurements along infinitesimally close axes. Generically, any three axes not in a plane will do:  
\be
\hat{\rho} \equiv \ket{\psi} \bra{\psi} 
      \Leftrightarrow \{ p_m^{(k)}, -s \leq m \leq s, k=1,2,3 \} \, ,
\label{purerho}
\ee
where the index $k$ is associated here with three unit vectors $\vec{e}_k$ such that 
$\vec{e}_1 \cdot ( \vec{e}_2 \times \vec{e}_3) \neq 0$. As before, this amounts to measuring more expectation values than there are free parameters in the pure state but it appears to be impossible to improve the result if one insists on using a Stern-Gerlach apparatus. 

The methods alluded to above are completely different from a mathematical point of view. As a matter of fact, it is much simpler to derive the result for two infinitesimally close axes than for the general case.  It does not seem possible to extend the argument which holds for infinitesimally close axes to a situation with finite angles between the axes. 

It is worth while to rephrase the problem solved here in algebraic form \cite{amiet+98b}. Consider  a faithful $(2s+1)$ dimensional representation of the group $SU(2)$ acting on the Hilbert space ${\cal H}_s$, and a fixed normalized (generic) state $\ket{\psi}$. Then, the following statement holds: 
\begin{equation}
e^{if({\hat s}_x)} \ket{\psi}    
  = e^{ig({\hat s}_y)} \ket{\psi}
  = e^{ih({\hat s}_z)}  \ket{\psi} 
\Leftrightarrow 
\hat f = \hat g = \hat h = \mbox{ const } \in {\sf I} \!\! {\sf R} \, ,
\label{algebraicf}
\end{equation}
where the functions $f(x), g(x)$ and $h(x)$ are polynomials of degree $2s$, $f(x)$$=$$ \sum_{\sigma =0}^{2s} f_\sigma x^\sigma$, etc. For simplicity, the axes in (\ref{algebraicf}) are associated with an orthogonal set of unit vectors $\vec{e}_k$. If there is a Pauli partner $\ket{{\widetilde \psi}}$ of the original state $\ket{\psi}$ at all, it must have the form given by the left-hand-side of (\ref{algebraicf}). For ${\vec e}_z$, e.g., one has 
\be
\ket{{\widetilde \psi}} 
         = e^{ih({\hat s}_z)} \ket{\psi} 
         = \sum_{m_z=-s}^{s} \ket{m_z} e^{ih(m_z)} \bra{m_z} \psi \rangle \, ,
\label{zphases}
\ee
which implies 
\be
|  \bra{m_z} {\widetilde \psi} \rangle |^2 = |  e^{ih(m_z)} \bra{m_z} \psi \rangle |^2 
             = |  \bra{m_z} \psi \rangle |^2 \, ,
\label{identicalmoduli}
\ee
and similarly for the two remaining directions. Thus, the states $\ket{{\widetilde \psi}}$ and $\ket{\psi}$ give rise to the {\em same} intensities along the three axes of quantization. The only consistent and nontrivial choice of the operators $\hat f, \hat g,$ and $\hat h$ indicated on the right-hand-side of (\ref{algebraicf}) leads, however, to a state $\ket{{\widetilde \psi}}$ which is just a multiple of $\ket{\psi}$ -- hence, $\ket{\psi}$ does not have a Pauli partner.      
\subsection*{Indirect measurement of arbitrary operators}
All successful schemes of state reconstruction, be it for particle or spin systems, provide ``half'' an answer to the problem to relate self-adjoint operators and observables in quantum mechanics. One is tempted to claim that all hermitean operators could be promoted to observables \cite{dirac58}: 
\begin{quotation}
In practice it may be very awkward or perhaps even beyond the ingenuity of the experimenter to devise an apparatus which could measure some particular operator, but the theory always allows one to imagine that the measurement can be made. 
\end{quotation}
Nevertheless, one usually reserves the notion ``observable'' for those operators which are known to be measured by some well-defined experimental setup. This requires an apparatus which, upon measuring ${\widehat{\cal O}}$, ``projects'' the original state into an eigenstate of the measured operator. On the basis of repeated measurements one 
would then be able to determine the expectation value of ${\widehat{\cal O}}$, to be compared with the calculated value $\mbox{Tr} ( \hat{\rho} \widehat{\cal O} )$. Without an apparatus measuring the operator ${\widehat{\cal O}}$, its expectation values seem to be out of reach. This state of affairs is judged as unsatisfactory \cite{wigner75}:
\begin{quotation}
There is, however, no rule which would tell us which self-adjoint operators are truly observables, nor is there any prescription known how the measurements are to be carried out, what apparatus to use, etc. In a theory with a positivistic undertone, this is a serious gap. 
\end{quotation}

With a working scheme of state reconstruction one can determine indirectly the expectation value of {\em any} operator ${\widehat{\cal O}}$ {\em without} measuring it. All one has to do is to reconstruct the (pure or mixed) state of the system at hand. This provides a parametrization of its density matrix in terms of a well-defined quorum ${\cal Q}$ of experimentally accessible expectation values $\langle {\widehat{ \cal O}}_q \rangle$, 
\be
\hat \rho = {\hat \rho} \left( \left\{ \langle {\widehat{ \cal O}}_q \rangle \right\} \right) \, .
\label{rhoinexp}
\ee
Subsequently, it is straightforward to {\em calculate} the expection value of ${\widehat{\cal O}}$ according to the rules of quantum mechanics giving
\be
\langle {\widehat{\cal O}} \rangle_{\hat \rho} 
         = \mbox{Tr} \left( \widehat{\cal O}_a 
                {\hat \rho} \left( \left\{ \langle {\widehat{ \cal O}}_q \rangle \right\} \right) \right)
         = \langle {\widehat{\cal O}} \rangle_{\hat \rho} 
            \left(  \left\{ \langle {\widehat{\cal O}}_q \rangle \right\} \right) \, .
\label{operatorexp}
\ee
Thus, the expectation values of operators without practicable measuring apparatus are  accessible experimentally -- the operators do not even have to be hermitean. 
From this point of view, one might consider the Stern-Gerlach apparatus as a {\em universal measuring device} for spin systems: when complemented with a scheme of state reconstruction, it enables one to extract all expectation values one can ever dream of.  

This argument can be turned around in order to ``test'' the predictions of quantum mechanics \cite{hegerfeldt98}. The 
expectation value of an observable with known measuring device can now be determined in {\em two} independent ways, using either the apparatus or the indirect method with a quorum ${\cal Q}$ of measurable operators. Comparison of the results checks the consistency of quantum mechanics. In a spin system, for example, this test is easily realized by looking at 
the expectation value of a spin component along some axis {\em not} involved in the quorum.
\subsection*{Quantum mechanics in terms of expectation values}
Each scheme of state reconstruction provides a new way  to represent the time evolution of a quantum system. Suggestively, one can write for a pure state\footnote{The restriction to pure states is not necessary; the argument to follow applies to mixed states as well.} 
\be
\ket{\psi} \Leftrightarrow \ket{ \{ \langle {\widehat{\cal O}}_q \rangle \} } \, , \quad
{\widehat{\cal O}}_q \in {\cal Q} \, ,
\label{suggestive}
\ee
where the operators ${\widehat{\cal O}}_q $ are assumed to provide a quorum ${\cal Q}$ for the system at hand. Effectively, one thinks here of the state $\ket{\psi}$ as parametrized by expectation values. The time evolution of the system is governed by Schr\"odinger's equation
\be
i\hbar \frac{d}{dt} \ket{\psi} = \widehat{H} \, \ket{\psi} \, ,
\label{schroe}
\ee
which transports $\ket{\psi(t_0)}$ at time $t_0$ along a path in Hilbert space to $\ket{\psi(t)}$ at some later time $t$. In view of the one-to-one relation (\ref{suggestive}), it is obvious that the traversed path has an unambiguous image in 
the {\em space of expectation values}. Therefore, a closed set of equations of motion 
for the elements of the quorum must exist,
\be
\frac{d}{dt} {\widehat{\cal O}}_q 
= f_{\widehat{H}}^{(q)} 
      \left( \{ \langle {\widehat{\cal O}}_q \rangle \}  \right) \, ,
\label{newdynamics}
\ee
where the function $f_{\widehat{H}}^{(q)}$ generates the correct transformation of the expectation values. It depends in a subtle way on both the quorum and the Hamiltonian $\widehat{H}$, and its detailed properties remain to be explored. 

This ``expectation-value representation'' of quantum mechanics is equivalent to any other representation. The time evolution of the system is represented as the motion of a point on a manifold in a space with axes corresponding to expectation values (or probabilities). This representation has the interesting property that it refers to measurable quantities only (in the form of expectation values): the wave function completely drops out. 
Conceptually, this approach differs truly from other formulations of quantum mechanics `without wave function' such as the phase-space representation in terms of Wigner functions, be it for particles \cite{baker58} or spin \cite{wooters87}. The novelty of the representation introduced here is due to expressing the quantal dynamics in terms of {\em directly} observable quantities.  
\subsection*{Conclusion}
State reconstruction for mixed and pure spin states has been reviewed. In both cases, measurements performed with a Stern-Gerlach apparatus are sufficient for a realistic quorum. Any working scheme of reconstruction  has two interesting consequences. On the one 
hand, it becomes possible to access experimentally the expectation values of arbitrary operators. This provides a partial solution to the problem of associating self-adjoint operators with observables: even if no measuring device for some operator ${\widehat{\cal O}} $ is known, one can determine its expectation value. On the other hand, new representations of the quantum dynamics follow from reliable reconstruction methods. In this framework, the time evolution of a quantum state corresponds to the trajectory of a point in a space of expectation values. Such a representation of quantum mechanics has 
the noteworthy feature to eliminate all unobservable elements from the theory. 
This provides a new and unexpected way to draw consequences of Schr\"odinger's idea \cite{schroedinger35} to think of the wave function as a ``Katalog der Erwartung.''  
\subsubsection*{Acknowledgements}
Financial support from the Swiss National Science Foundation is gratefully acknowledged.
\end{document}